\documentclass[pra,twocolumn, notitlepage,superscriptaddress, nofootinbib]{revtex4-1}

\usepackage{fancyhdr}
\fancyheadoffset[L,R]{\headrulewidth}
\renewcommand{\headrulewidth}{0.6pt}

\usepackage{cancel}
\usepackage[normalem]{ulem}

\usepackage{cancel}

\usepackage{graphicx}
\usepackage{tikz,pgf}
\usepackage{color}
\usepackage{array}
\usepackage{xcolor}
\usepackage[dvips]{epsfig}
\usepackage{epsfig}
\usepackage{enumerate}
\usepackage[latin1]{inputenc}
\usepackage[T1]{fontenc}
\usepackage[hang]{subfigure}
\usepackage{bbm, dsfont}
\usepackage{amsfonts,amsmath,amssymb,stmaryrd}
\usepackage{ulem}

\usepackage{bbold}

\usepackage{simplewick}

\newcommand{\ga}{\ga}

\newcommand{\bl}{\begin{linenomath*}}
\newcommand{\el}{\end{linenomath*}}

\newcommand{\bea}{\begin{eqnarray}}
\newcommand{\eea}{\end{eqnarray}}

\renewcommand{\ga}{\hat\gamma}

\definecolor{dgreen}{rgb}{0.0, 0.5, 0.0}

\usepackage{hyperref}

\begin{document}

\title{Libration of strongly-oriented polar molecules inside a superfluid}

\author{E.~S.~Redchenko}
\affiliation{IST Austria (Institute of Science and Technology Austria), Am Campus 1, 3400 Klosterneuburg, Austria}

\author{Mikhail Lemeshko}
\email{mikhail.lemeshko@ist.ac.at}
\affiliation{IST Austria (Institute of Science and Technology Austria), Am Campus 1, 3400 Klosterneuburg, Austria}
\affiliation{Kavli Institute for Theoretical Physics, University of California, Santa Barbara, CA 93106, USA}

\begin{abstract}

We study a polar molecule immersed into a superfluid environment, such as a helium nanodroplet or a Bose-Einstein condensate, in the presence of an intense electrostatic field. We show that coupling of the molecular pendular motion, induced by the field, to the fluctuating bath leads to formation of \textit{pendulons} -- spherical harmonic librators dressed by a field of many-particle excitations. We study the behavior of the pendulon in a broad range of molecule-bath and molecule-field interaction strengths, and reveal that its spectrum features series of instabilities which are absent in the field-free case of the angulon quasiparticle. Furthermore, we show that an external field allows to fine-tune the positions of these instabilities in the molecular rotational spectrum. This opens the door to detailed experimental studies of redistribution of orbital angular momentum in many-particle systems. 

\end{abstract}

\maketitle

\section{Introduction}

Recently, it was shown that molecular rotation inside a superfluid leads to formation of a quasiparticle of a new kind -- the \textit{angulon}. The angulon represents a quantum rotor dressed by a field of many-particle excitations~\cite{SchmidtLem15, SchmidtLem16, LemSchmidtChapter}, and in principle can be thought of as a rotational analogue of the well-studied polaron quasiparticle~\cite{AppelPolarons, EminPolarons, PolaronsExcitons, polaron_review1, Devreese15}.  It was demonstrated, however, that the non-Abelian algebra used to describe quantum rotations, as well as the discrete spectrum of the rotor's eigenvalues, render the angulon physics substantially different from that of any other impurity problem~\cite{SchmidtLem15, SchmidtLem16, LemSchmidtChapter}.

For instance, once placed inside superfluid, the molecular states become `dressed' by a cloud of virtual phonon excitations -- the process which lowers the molecular energy. While this effect also takes place for polarons, in the case of molecules the magnitude of the shift depends on the rotational state the molecule is in. This results in renormalization of the molecular rotational constant -- the effect previously observed for several molecules inside superfluid helium nanodroplets~\cite{ToenniesVilesov04}. The angulon theory allows for a simple interpretation of such a renormalization in terms of the rotational Lamb shift~\cite{SchmidtLem15, SchmidtLem16, LemSchmidtChapter}, which is an exact phononic analogue of the photonic Lamb shift of quantum electrodynamics~\cite{Karshenboim2005}. It was shown that even in the context of weakly-interacting ultracold gases, the magnitude of this shift is large enough to be accessible in modern experiments~\cite{Bikash16}.

Moreover, interaction of a molecular impurity with a superfluid leads to appearance of the many-body-induced fine structure in the angulon rotational spectrum. The latter occurs due to resonant exchange of angular momentum between the impurity and the many-body bath, and has no direct analogue in isolated isolated atoms or molecules.

In this contribution, we study the behavior of the angulon in the presence of an intense electrostatic field. We show that in the strong-field limit the angulon turns into the `pendulon' -- a quantum spherical librator, whose pendular motion is altered by the field of phonon excitations. The paper is organized as follows. In Sec.~\ref{sec:Hamil} we describe the angulon Hamiltonian in the presence of an external electrostatic field. Next, in Sec.~\ref{sec:pendulon} we consider the strong-field limit of this Hamiltonian, which  describes the pendulon quasiparticle. In Sec.~\ref{sec:variational} we present a variational approach to the pendulon Hamiltonian and show that it is possible to reformulate it in a diagrammatic language, thereby acquiring access to the entire spectrum of the system. Next, in Sec.~\ref{sec:spectrum} we study how the pendulon spectra change depending on the superfluid density and external electric field, and uncover series of instabilities accompanied by the transfer of angular momentum from the impurity to the bath. Finally, Sec.~\ref{sec:conclusions} outlines the conclusions of this work.

\section{The angulon Hamiltonian}

\label{sec:Hamil}

We consider a linear rotor molecule immersed into a superfluid, as described by the angulon Hamiltonian~\cite{SchmidtLem15}, in the presence of an external electrostatic field:
\begin{equation}
\label{eq:Ham1}
\hat H = \hat H_\text{m-f} +  \hat H_\text{b} +  \hat H_\text{m-b}
\end{equation}
Here the first term,
\begin{equation}
\label{eq:Hmf}
 \hat H_\text{m-f} = B\hat{\bf{J}}^2 - d \mathcal{E}\cos \hat \theta,
\end{equation}
describes a polar molecule with a dipole moment $d$  in an electrostatic field of magnitude $\mathcal{E}$, which defines the laboratory-frame $z$-axis. For simplicity, we consider the case of a linear-rotor impurity, whose low-energy spectrum is characterized by one rotational constant, $B$. The molecular eigenstates, $\vert j,m\rangle$, are labeled by the angular momentum, $j$, and its projection, $m$,  onto the $z$-axis. The theory, however, is straightforward to extent to more complex species, such as symmetric and asymmetric tops~\cite{TownesSchawlow, BernathBook}. For convenience, we will express the strength of the molecule-field interaction using a dimensionless parameter,
\begin{equation}
\label{eq:eta}
\eta \equiv \frac{d \mathcal{E}}{B} 
\end{equation}

The boson kinetic energy is given by the second term of Eq.~\eqref{eq:Ham1},
\begin{equation}
\label{eq:Hb}
 \hat H_\text{b} = \sum_{k\lambda\mu}\omega_k\hat{b}^+_{k\lambda\mu}\hat{b}_{k\lambda\mu},
\end{equation}
where $\omega_k$ is the dispersion relation. Here and below we use the labelling $\sum_{k}\equiv\int dk$ and set $\hbar\equiv1$.  Furthermore, in Eq.~\eqref{eq:Hb} the bosonic creation and annihilation operators, $\hat{b}^+_{\bf{k}}$ and $\hat{b}_{\bf{k}}$,  are expressed in the angular momentum basis, $\hat{b}^+_{k\lambda\mu}=k(2\pi)^{-3/2}\int d\Omega_k\hat{b}^+_\mathbf{k}i^\lambda Y^*_{\lambda\mu}(\Omega_k)$. Here, $k=|\bf{k}|$, while $\lambda$ and $\mu$ define the boson angular momentum and its projection onto the $z$-axis, see Refs.~\cite{SchmidtLem15, SchmidtLem16, LemSchmidtChapter} for details. 

Finally, the third term of Eq.~\eqref{eq:Ham1}
 gives the molecule-boson interactions,
\begin{equation}
\label{eq:Hmb}
 \hat  H_\text{m-b} =  \sum_{k\lambda\mu}U_{\lambda}(k)\left[Y^*_{\lambda\mu}(\hat{\theta},\hat{\phi})\hat{b}^+_{k\lambda\mu} + Y_{\lambda\mu}(\hat{\theta},\hat{\phi})\hat{b}_{k\lambda\mu}\right]
\end{equation}
Note that the interaction  depends explicitly on the molecular orientation in the laboratory frame, as given by the spherical harmonic operators, $Y_{\lambda\mu}(\hat{\theta},\hat{\phi})$~\cite{Varshalovich}. The magnitude of the coupling to phonons carrying angular momentum $\lambda$ and linear momentum $k$ is parametrized by $U_{\lambda}(k)$. Originally, Eq.~\eqref{eq:Hmb}
 was derived to describe an ultracold molecule coupled to a weakly-interacting Bose-Einstein condensate (BEC), which leads to an analytic expression for $U_{\lambda}(k)$~\cite{SchmidtLem15},
\begin{equation}
\label{Ulamk}
	U_\lambda(k) =   \left[\frac{8 n k^2\epsilon_k}{\omega_k(2\lambda+1)}\right]^{1/2} \int dr r^2 V_\lambda(r) j_\lambda (kr)
\end{equation}
The latter assumes that in the impurity frame, $\{x',y',z' \}$, the  interaction between the linear rotor and  a bosonic atom is expanded as
\begin{equation}
\label{VimpBos}
	V_\text{imp-bos} (\mathbf{r'}) = \sum_\lambda   V_\lambda (r') Y_{\lambda 0} ( \Theta', \Phi' ),
\end{equation}
with $V_\lambda(r')$ giving the interaction potential in the corresponding  angular momentum channel. The prefactor of Eq. \eqref{Ulamk} depends on the bath number density, $n$, the kinetic energy of the bare atoms, $\epsilon_k$, and the bath dispersion relation, $\omega_k$.

While for more involved cases, such as molecules in superfluid helium nanodroplets, Eq.~\eqref{eq:Ham1}
 is not exact, it can still be looked at from a phenomenological point of view, with the coupling constants $U_{\lambda}(k)$ and dispersion relation $\omega_k$ extracted from experiment or \textit{ab initio} calculations. Therefore, in what follows we will approach the Hamiltonian ~\ref{eq:Ham1} 
 from a completely general perspective.

\begin{figure}[t]
\centering
\includegraphics[width=0.7\linewidth]{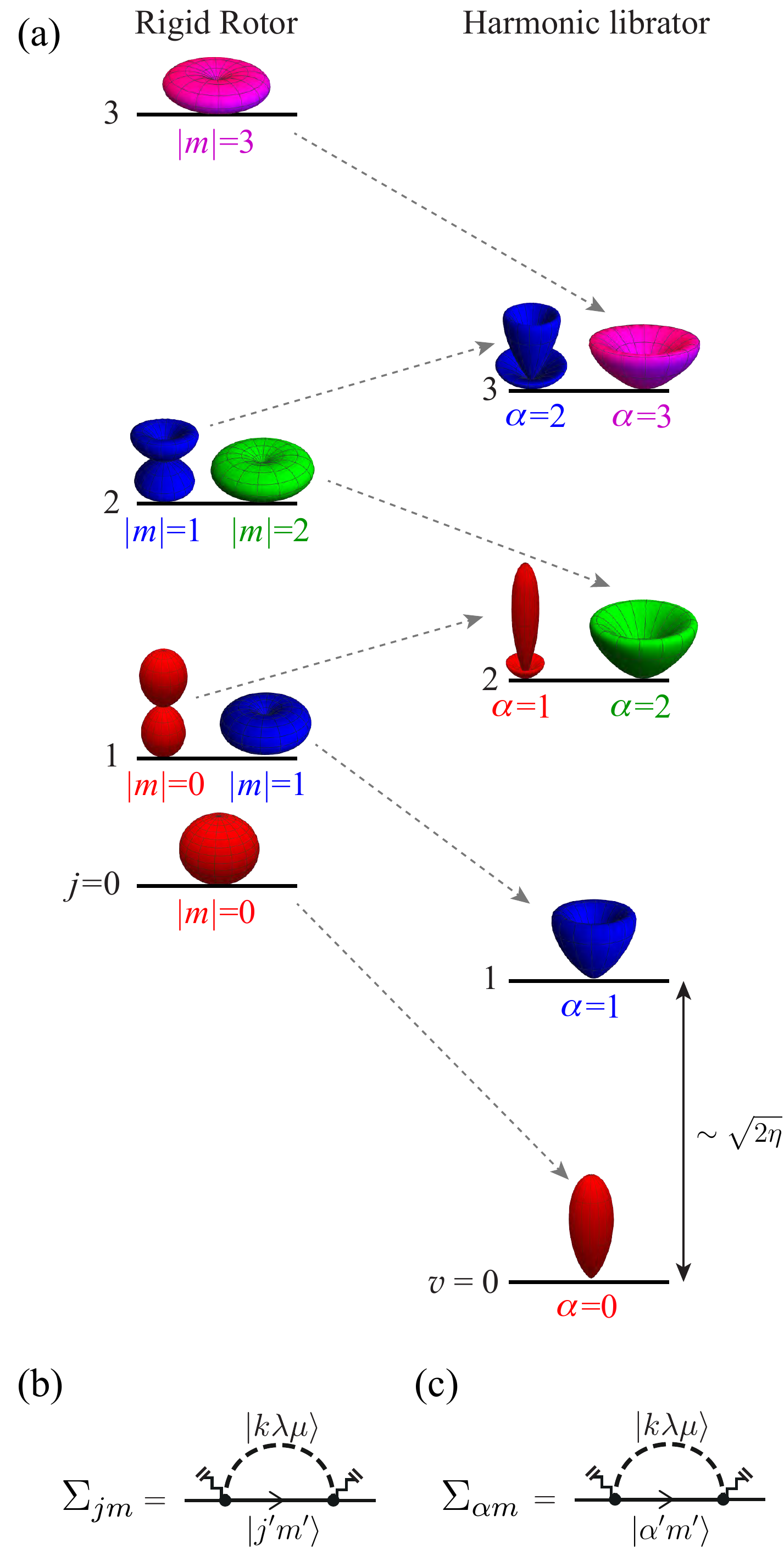}
\caption{(a) Correlation diagram illustrating the crossover between the molecular rotational states in zero field (left) and in an intense electrostatic field (right). The $m$ quantum number is conserved in the presence of a field. (b), (c) Diagrammatic representation of the corresponding self-energies in the presence of a superfluid, Eq.~\eqref{eq:Selfenergy}.}
\label{fig:levels}
\end{figure}

\section{Angulon's strong-field limit: the pendulon}
\label{sec:pendulon}

Let us focus on the strong-field regime, $\eta \gg 1$, which has been thoroughly studied for molecules in the gas phase in the past \cite{RostPRL92, FriedrichCanJPhys94, LemPRA11, LemNJP11, LemKreDoyKais13}. Most importantly, it has been shown that in such a regime, the molecule-field Hamiltonian $ \hat H_\text{m-f}$ can be reduced to that for a spherical harmonic librator. The latter, in turn, can be diagonalized exactly leading to the following eigenvalues:
\begin{equation}
\label{eq:Eam}
	E_{\alpha m} = \sqrt{2 \eta} (v +1) - \eta,
\end{equation}
where the vibrational quantum number $v   = 2 \alpha - |m| = 0, 1, 2, \dots $.

The eigenstates of the harmonic librator in the angular representation are given by:
\begin{multline}
\label{eq:Psiam}
	\langle \theta, \phi \vert  \alpha m \rangle =  N_{\alpha m} \theta^{|m|+1/2} L_{\alpha-|m|}^{|m|} \left (\frac{\sqrt{2 \eta} \theta^2}{2} \right)  \sin^{-\frac{1}{2}}\theta  \\
	\times \exp \left(-\frac{\sqrt{2 \eta} \theta^2}{4} + i m \phi \right),
\end{multline}
where $L_n^k (x)$ are associated Laguerre polynomials and the normalization constant is given by:
 \begin{equation}
\label{eq:Nam}
	N_{\alpha m } = \left( \frac{\sqrt{2 \eta}}{2} \right)^{\frac{|m|+1}{2}} \left[ \frac{2 (\alpha - |m|)!}{\alpha!} \right]^{\frac{1}{2}}
\end{equation}
The projection of molecular angular momentum on the laboratory-frame $z$ axis, $|m| = 0,1, 2, \dots, \alpha$, gives the number of $\phi$ nodes of the wavefunction. It is a good quantum number in the presence of a linearly polarized field. The total number of the $\theta$ nodes, on the other hand, is given by $n_\theta = \alpha - |m|$. Thus, although the state \ref{eq:Psiam} represents a superposition of several field-free rotational states,
\begin{equation}
\label{eq:PsiamViaY}
	 \vert  \alpha m \rangle =  \sum_{j \geq |m|} g_{jm}^{(\alpha)} \vert j m \rangle
\end{equation}
the quantum number $\alpha$ adiabatically turns into $j$ in the limit of $\eta \to 0$. Fig.~\ref{fig:levels}(a) illustrates the crossover between the field-free states of a rigid rotor and the harmonic librator states of Eq.~\eqref{eq:Eam}, with the corresponding change of the eigenfunctions. One can see that the states of harmonic librator are strongly-oriented with respect to the $z$-axis.

Thus, in the limit of an intense electrostatic field, the Hamiltonian~\ref{eq:Ham1} can be rewritten as:
\begin{equation}
\label{eq:Ham2}
\hat H_\text{pend} =  \sum_{\alpha m} E_{\alpha m} |\alpha m\rangle\langle\alpha m|+  \hat H_\text{b} +  \hat H_\text{m-b}
\end{equation}
Hereafter we will refer to Eq.~\eqref{eq:Ham2} as the `pendulon Hamiltonian.'

\section{Variational solutions}
\label{sec:variational}

By analogy with Ref.~\cite{SchmidtLem15}, we adapt a variational ansatz based on single-phonon excitations:
\begin{equation}
\label{eq:VarWF}
	 |\psi_{\alpha m}\rangle  =  Z_{\alpha m}^{1/2} |0\rangle|\alpha m\rangle + \sum_{k \lambda \mu \alpha' m'} \beta_{k \lambda \mu}^{(\alpha')}b_{k \lambda \mu}^{\dagger} |0\rangle|\alpha' m'\rangle
\end{equation}
Note that no Clebsch-Gordan coefficient appears in Eq.~\eqref{eq:VarWF} since the total angular momentum is not a good quantum number in the presence of a field. Selection rule on projections, $m' \equiv m-\mu$, comes out naturally from the variational procedure outlined below.

\begin{figure}[b]
\centering
\includegraphics[width=0.9\linewidth]{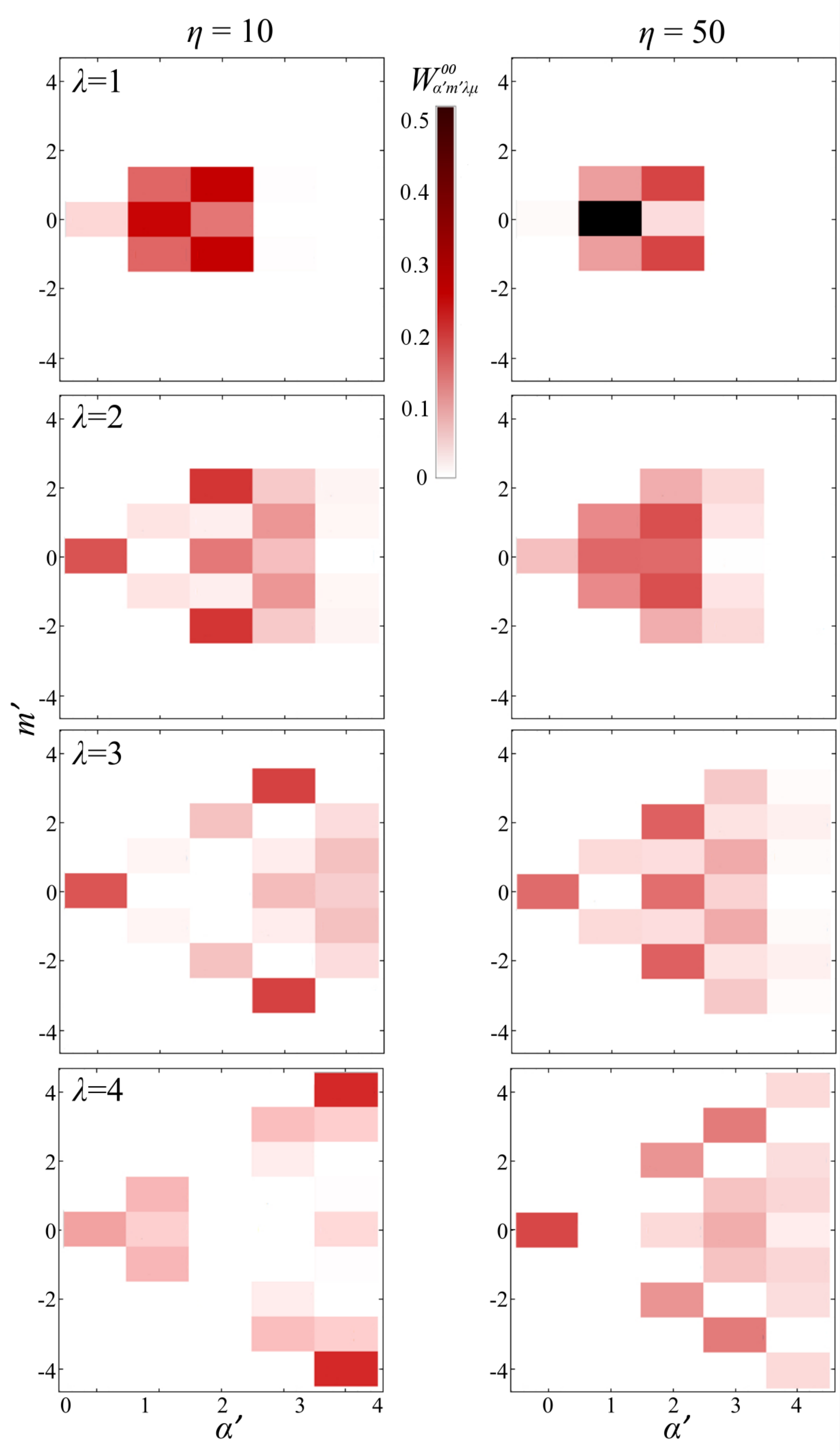}
\caption{Selection rules on virtual phonon excitations for the ground pendulon state, $|\psi_{0 0}\rangle$, as given by the magnitude of the coupling coefficients, Eq.~\eqref{eq:W}.}
\label{fig:Wcoef}
\end{figure}

Using the ansatz~\ref{eq:VarWF}, we minimize the energy, $E=\langle\psi_{\alpha m}| \hat H_\text{pend} |\psi_{\alpha m}\rangle\langle\psi_{\alpha m}|\psi_{\alpha m} \rangle $, or, equivalently, the functional $F = \langle\psi_{\alpha m}| \hat H_\text{pend} - E |\psi_{\alpha m}\rangle$.
In Refs.~\cite{SchmidtLem15, SchmidtLem16, LemSchmidtChapter} it was shown that for a variational state of the form~\ref{eq:VarWF}, the resulting equations can be cast in form of the Dyson equation:
\begin{equation}
\label{eq:Dyson}
	[G_{\alpha m} (E)]^{-1} \equiv [G_{\alpha m}^{(0)} (E)]^{-1} - \Sigma_{\alpha m} (E) = 0
\end{equation}
Here $[G_{\alpha m} (E)]^{-1}$ is the total pendulon Green's function and
\begin{equation}
\label{eq:G0}
	G_{\alpha m}^{(0)} (E) = \frac{1}{E_{\alpha m} - E}
\end{equation}
is the `bath-free' Green's function of a molecule in an electric field. The pendulon self-energy is given by:
\begin{equation}
\label{eq:Selfenergy}
\Sigma_{\alpha m} (E) =  \sum_{\substack{k \lambda \mu \\ \alpha' m'}} \frac{2 \lambda+1}{4 \pi} \frac{U_\lambda (k)^2~W_{\alpha' m', \lambda \mu}^{\alpha m}}{E_{\alpha' m'} - E+ \omega_k}
\end{equation}
where
\begin{equation}
\label{eq:W}
W_{\alpha' m', \lambda \mu}^{\alpha m} = \left| \sum_{j j'} C_{j0, \lambda 0}^{j' 0}~C_{j' m', \lambda \mu}^{j m}~g_{jm}^{(\alpha)}~g_{j' m'}^{(\alpha')} \right|^2
\end{equation}
with $C_{j_1 m_1, j_2 m_2}^{j_3 m_3}$ the Clebsch-Gordan coefficients~\cite{Varshalovich}. One can see that for the zero-field case, $g_{jm}^{(\alpha)} \equiv \delta_{\alpha, j}$, Eqs.~\ref{eq:Selfenergy} and \eqref{eq:W} reduce to the previously derived self-energy of the free angulon~\cite{SchmidtLem15}. The Feynman-diagram representation of the self-energies of the angulon and pendulon are shown in Figs.~\ref{fig:levels}(b) and (c), respectively.

The  coefficients $W_{\alpha' m', \lambda \mu}^{\alpha m}$ determine the selection rules on the virtual phonon excitations shown in Figs.~\ref{fig:levels}(c), and thereby provide a strong-field analogue of the Clebsch-Gordan coefficient $(C_{j0, \lambda 0}^{j' 0})^2$ appearing in the field-free case. As an example, Fig.~\ref{fig:Wcoef} shows the  $W_{\alpha' m', \lambda \mu}^{0 0}$ coefficients which determine the self-energy of the ground pendulon state. One can see that the distribution of the dominant transitions (dark red) as well as forbidden ones (white) changes depending on the field strength, $\eta$, and the symmetry of the impurity-bath coupling, $\lambda$.

Thus, having the pendulon Green's function, Eq.~\eqref{eq:Dyson}, at hand,  we acquire access not only to the ground-state properties but also to the entire excitation spectrum of the system. The latter is given by the spectral function~\cite{AltlandSimons},
 \begin{equation}
\label{SpecFunc}
 \mathcal A_{\alpha m} (E)= \text{Im}[G_{\alpha m} (E+i0^+)],
 \end{equation}
which in addition provides insight  into the quasiparticle properties of the angulon. 

\section{The pendulon spectrum}
\label{sec:spectrum}

In order to elucidate the pendulon's properties, here we provide numerical results on the spectral function, Eq.~\eqref{SpecFunc}. As an example, we consider a molecule immersed into a weakly-interacting BEC, whose dispersion is given by the Bogoliubov relation, $\omega_k=\sqrt{\epsilon_k(\epsilon_k+2 g_{\text{bb}} n)}$, where $g_\text{bb}=4\pi a_{\text{bb}}/m$ \cite{Pitaevskii2003} with $a_\text{bb}>0$ the boson-boson scattering length and $m$ the boson mass; $\epsilon_k = \hbar^2 k^2/(2m)$ gives the boson kinetic energy.

 Furthermore, in what follows, we adapt the dimensionless units where energy is measured in units of $B$ and  distance in units of $(m B)^{-1/2}$ (in addition to $\hbar \equiv 1$ as defined above). Thus the unit of the bath number density, $(m B)^{3/2}$,  depends on the particular molecule and bosons one considers.

A typical molecule-atom potential is usually dominated by first few angular-momentum components, $V_\lambda (r')$ of Eq.~\eqref{VimpBos}; furthermore, in several cases, the $\lambda=0, 2$ components provide the main contribution~\cite{StoneBook13, Bikash16}. Therefore, as a qualitative model, we adapt the following Gaussian potentials:
 \begin{equation}
\label{Vlam}
V_\lambda (r') = u_\lambda (2\pi)^{-3/2} e^{-r^2/(2r_\lambda^2)}
 \end{equation}
 and assume the interactions present only in $\lambda=0, 2$ channels. Taking into account a typical shape of the molecule-atom potentials~\cite{StoneBook13, Bikash16}, we choose the magnitudes of the interactions and their corresponding ranges as  $u_0 = 1.75 \, u_2 = 218$ and  $r_0 = r_2 = 1.5$. We set the boson-boson scattering length $a_\text{bb}=3.3$, which approximately reproduces the speed of sound in superfluid $^4\text{He}$  for a molecule with $B=2\pi \times 1~\text{GHz}$~\cite{DonnellyHe98}.

\begin{figure}[b]
\centering
\includegraphics[width=0.8\linewidth]{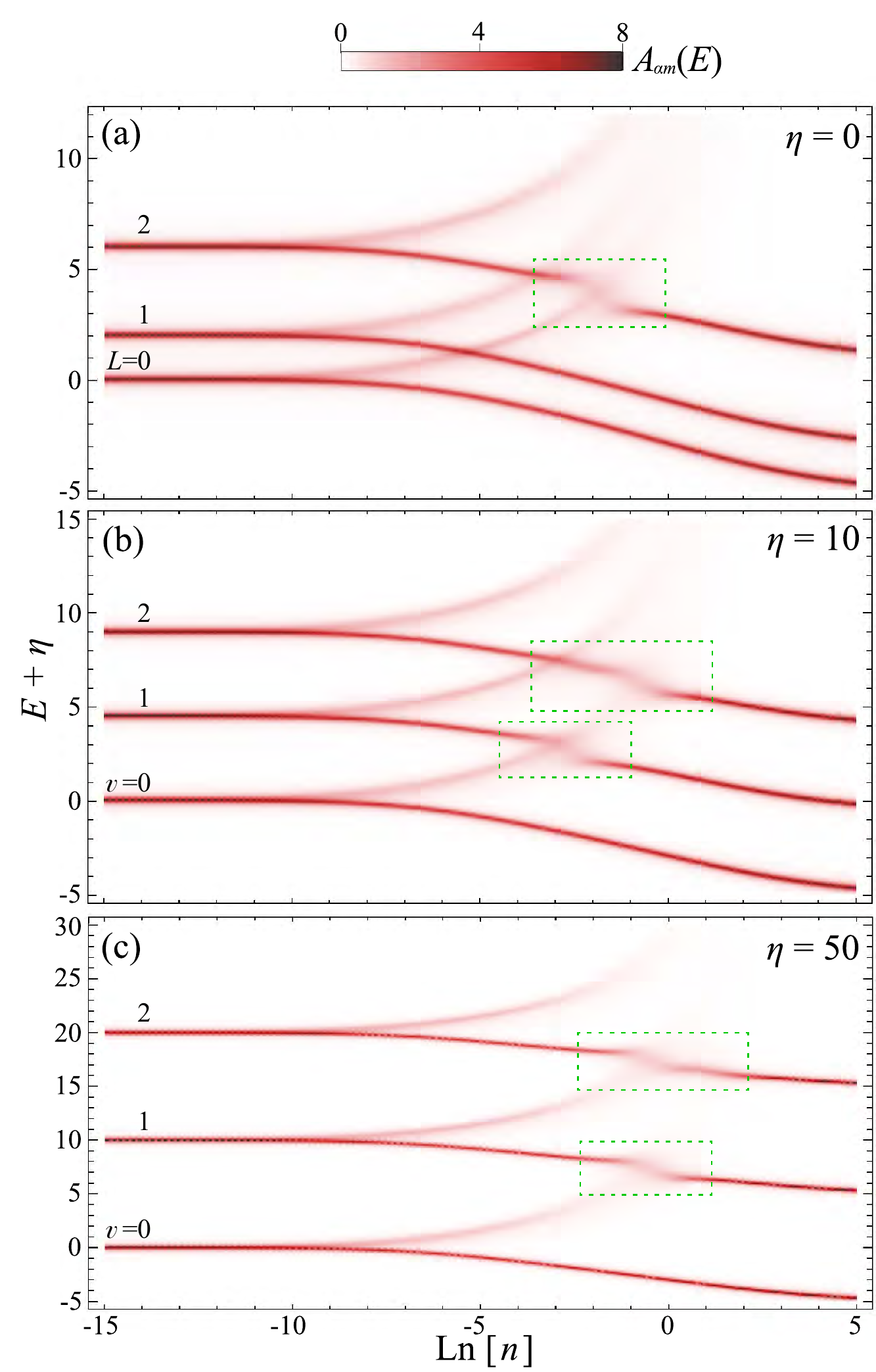}
\caption{Dependence of the pendulon spectral function, $A_L(E)$, on the dimensionless density of the superfluid, $n$, for selected electrostatic field intensities, $\eta$. Panel (a) corresponds to the field-free limit of the pendulon, i.e.\  the angulon quasiparticle~\cite{SchmidtLem15}.}
\label{fig:spec_n}
\end{figure}

Fig.~\ref{fig:spec_n} shows the dependence of the spectral function, Eq.~\eqref{SpecFunc}, on the superfluid density $n$, for various intensities of the electrostatic field, $\eta$. For a molecule with a dipole moment $d = 1$~Debye and $B=2\pi \times 1~\text{GHz}$, $\eta=1$ corresponds to a field of $\mathcal{E} \approx 2$~kV/cm, i.e.\ experimentally accessible fields correspond to $\eta \leq 50$.

Fig.~\ref{fig:spec_n}(a) presents the spectral function for the zero-field limit of the pendulon -- the angulon quasiparticle~\cite{SchmidtLem15, SchmidtLem16, LemSchmidtChapter, Bikash16}. In zero field, the only conserved quantity is total angular momentum of the system, $L$, and its projection, $M$, which for vanishing molecule-bath interactions coincide with the molecular quantum numbers, $j,m$. The many-body bath leads to several new features in the impurity spectrum. First, one can notice that around $\text{Ln}[n] = -7.5$ each $L$ level splits into two components, which was referred to as `many-body-induced fine structure of the first kind' in Ref.~\cite{SchmidtLem15}. The latter occurs due to the interaction between the state with no phonons, $\langle L=j| \langle 0|$, and a state with one spherically-symmetric phonon, $\langle L=j| \langle\lambda=0|$. Since the corresponding matrix element involves only the spherically-symmetric part of the molecule-bath interaction, $U_0 (k)$ of Eq.~\eqref{eq:Hmb}, the resulting effect does not change in the presence of an external field.

Furthermore, as it was previously shown, the angulon spectrum features instabilities associated with the resonant transfer of  angular momentum between the molecule and the superfluid, dubbed `many-body-induced fine structure of the second kind' in Ref.~\cite{SchmidtLem15}. The latter can take place once the   impurity-bath interactions are strong enough to push the angulon formed around the molecule in state $j$ close to the phonon continuum attached to the molecular state $j-1$. The emerging instability corresponds to the transfer of one unit of angular momentum from the molecule to the bath, induced by anisotropic ($\lambda>0$) molecule-bath coupling. The regions where such instabilities take place are highlighted in Fig.~\ref{fig:spec_n} by green dashed frames. For our choice of parameters, the angulon instabilities between neighbouring states are forbidden by the selection rules arising due to the symmetry of the impurity-bath interaction. Namely, the only anizotropic term present, $\lambda = 2$, couples the states with $j=j;~j\pm2$ and cannot induce any process changing the molecular angular momentum by one.\footnote{Note that such instabilities are allowed in the presence of $\lambda=1$ interactions, see Ref.~\cite{SchmidtLem15} for details.}  However, the instabilities between the states with $\Delta j = 2$ are still allowed, as one can see from the behavior of $A_{L=2}$ in the vicinity of $\text{Ln}[n] = -1$.

\begin{figure}[b]
\centering
\includegraphics[width=0.8\linewidth]{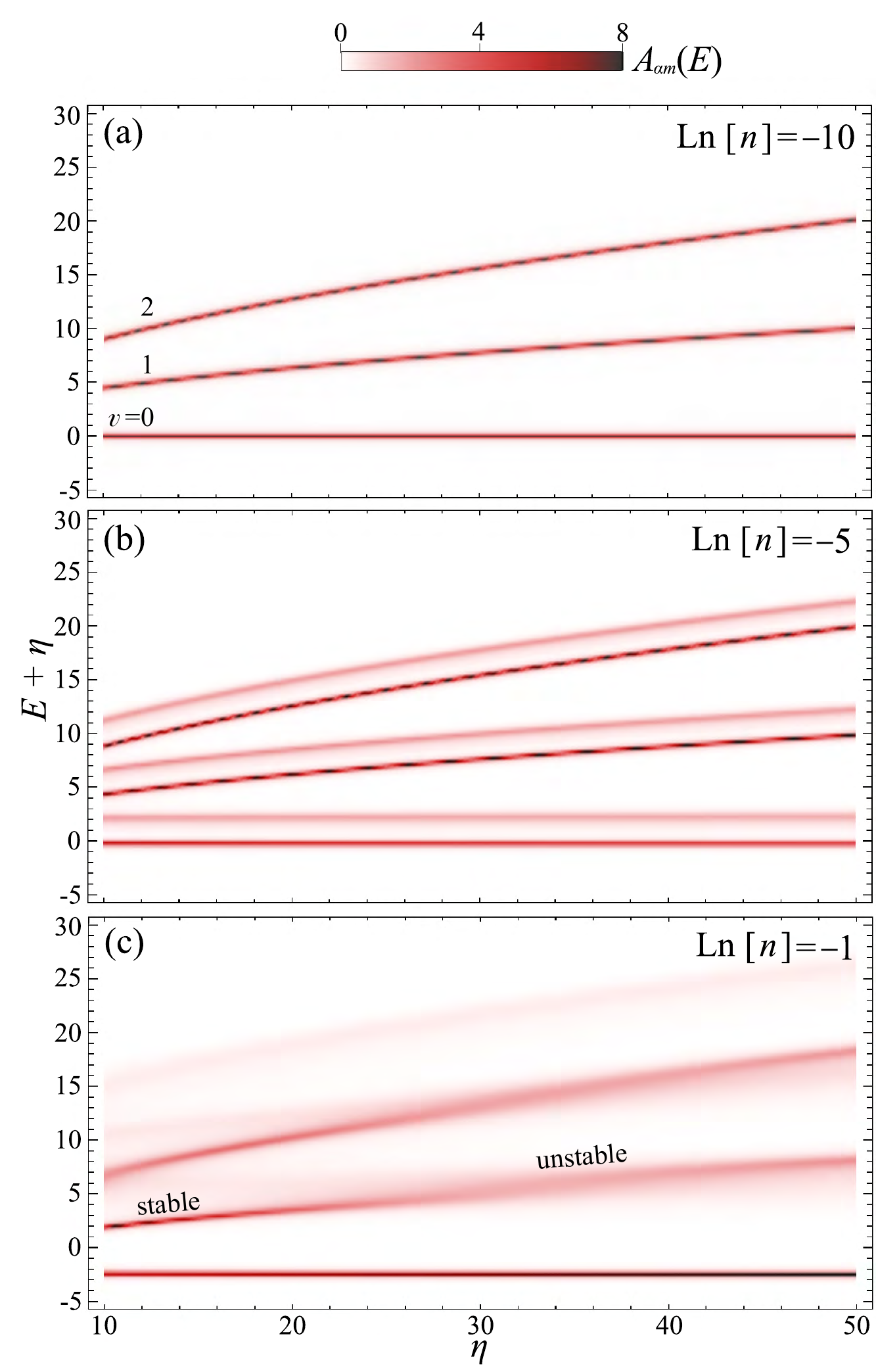}
\caption{Dependence of the pendulon spectral function, $A_L(E)$, on the intensity of the electrostatic field, $\eta$, for selected densities, $n$. Panel (c) shows the transition from stable to unstable behavior.}
\label{fig:spec_eta}
\end{figure}

Figs.~\ref{fig:spec_n}(b) and (c) show the pendulon spectra in moderate, $\eta=10$, and intense, $\eta=50$, electrostatic fields. In this case, the only conserved quantity of the system is the projection of the total angular momentum onto the field axis. Note  that the $v=2$ state of a molecule in a strong field is double-degenerate, however, its components behave qualitatively similar in the presence of a superfluid. The most peculiar effect due to the external field is the occurrence of the instabilities between the neighbouring pendulon states, which take place due to the relaxed selection rules on the angular momentum exchange. As a result, the pendulon spectrum features the instabilities independently of the exact symmetry of the anisotropic molecule-bath interactions, and thereby makes it easier to observe them in experiment. Furthermore, at $\eta=10$ there occur a `double instability' around $\text{Ln}[n] = -1$, which corresponds to the resonant transitions changing the molecular state by  $\Delta v = 1$ and $2$. Finally, one can see that the position of the instability shifts to larger densities $n$ at higher intensities of the electrostatic field.

In order to get a deeper insight into the field-dependence of the pendulon states, in Fig.~\ref{fig:spec_eta} we show the spectrum as a function of $\eta$, for a few representative superfluid densities. Fig.~\ref{fig:spec_eta}(a) corresponds to the pendulon states in the limit of low superfluid densities. Due to the weak interactions with the environment, the field-dependence is very close to that for a gas-phase molecule, i.e. the splitting between the neighboring states grows proportionally to $\sqrt{2 \eta}$. At higher densities, Fig.~\ref{fig:spec_eta}(b), one can see an additional metastable state attached to the stable pendulon level. As discussed above, it corresponds to dressing of the strongly-oriented molecular state with a spherically-symmetric phonon, $\lambda=0$. Due to the spherical symmetry, the position of this peak relatively to the stable state does not get altered by an external field. Finally, Fig.~\ref{fig:spec_eta}(c) illustrates a transition from the stable pendulon state (sharp peak) to an instability, accompanied by a transfer of the angular momentum to the superfluid environment (broad peak). Thus, using external fields allows to fine-tune the instabilities' positions thereby paving the way to studying the angular momentum transfer in many-body systems using molecules interacting with superfluids.

\section{Conclusions}
\label{sec:conclusions}

In this work we studied the properties of the pendulon quasiparticle, which represents an angulon in the presence of an intense electrostatic field. We have shown that an electric field relaxes the selection rules on the angular momentum exchange between the molecule and the bath, which results in a series of instabilities absent for the angulon. In other words, a field renders the instabilities universal, i.e.\ independent on the details of the molecule-boson potential energy surface. Furthermore, a field acts as an additional knob for altering the positions of the instabilities in the absorption spectrum, thereby opening the door for detailed experimental studies of redistribution of angular momentum in many-particle systems.

The most natural candidate for the experimental implementation of the scheme are molecules trapped inside superfluid helium nanodroplets~\cite{ToenniesVilesov04}, where external fields can be applied to confine molecular rotation~\cite{PentlehnerPRA13}, while the molecule-superfluid interactions are strong enough to modify the rotational spectrum of the impurity~\cite{MorrisonJPCA13}.

While the theory presented in this paper focuses on a closed-shell $^1\Sigma$ linear polar molecule, the model is straightforward to extend to more complex species rotating as symmetric and asymmetric tops~\cite{LevebvreBrionField2, TownesSchawlow, LemKreDoyKais13}. Due to a more involved rotational structure of such species, the pendulon spectrum is expected to be even richer than that described in the present study. Finally, it would be of great interest to investigate the field effect on weakly-bound halo-species (such as Feshbach molecules) immersed into a Bose-Einstein condensate~\cite{LemFriPRArapid09, LemFriPRL09, LemFriJPCA10}. There, novel effects are expected to take place due to a strong coupling of both molecular rotational and vibrational degrees of freedom to the condensate excitations.

\section{Acknowledgements}

We thank Klaus M{\o}lmer for suggesting the term `pendulon.' This project has received funding from the European Union's Horizon 2020 research and innovation programme under the Marie Sk{\l}odowska-Curie Grant Agreement No.~665385.

\bibliography{pendulon}

\begin{thebibliography}{29}%
\makeatletter
\providecommand \@ifxundefined [1]{%
 \@ifx{#1\undefined}
}%
\providecommand \@ifnum [1]{%
 \ifnum #1\expandafter \@firstoftwo
 \else \expandafter \@secondoftwo
 \fi
}%
\providecommand \@ifx [1]{%
 \ifx #1\expandafter \@firstoftwo
 \else \expandafter \@secondoftwo
 \fi
}%
\providecommand \natexlab [1]{#1}%
\providecommand \enquote  [1]{``#1''}%
\providecommand \bibnamefont  [1]{#1}%
\providecommand \bibfnamefont [1]{#1}%
\providecommand \citenamefont [1]{#1}%
\providecommand \href@noop [0]{\@secondoftwo}%
\providecommand \href [0]{\begingroup \@sanitize@url \@href}%
\providecommand \@href[1]{\@@startlink{#1}\@@href}%
\providecommand \@@href[1]{\endgroup#1\@@endlink}%
\providecommand \@sanitize@url [0]{\catcode `\\12\catcode `\$12\catcode
  `\&12\catcode `\#12\catcode `\^12\catcode `\_12\catcode `\%12\relax}%
\providecommand \@@startlink[1]{}%
\providecommand \@@endlink[0]{}%
\providecommand \url  [0]{\begingroup\@sanitize@url \@url }%
\providecommand \@url [1]{\endgroup\@href {#1}{\urlprefix }}%
\providecommand \urlprefix  [0]{URL }%
\providecommand \Eprint [0]{\href }%
\providecommand \doibase [0]{http://dx.doi.org/}%
\providecommand \selectlanguage [0]{\@gobble}%
\providecommand \bibinfo  [0]{\@secondoftwo}%
\providecommand \bibfield  [0]{\@secondoftwo}%
\providecommand \translation [1]{[#1]}%
\providecommand \BibitemOpen [0]{}%
\providecommand \bibitemStop [0]{}%
\providecommand \bibitemNoStop [0]{.\EOS\space}%
\providecommand \EOS [0]{\spacefactor3000\relax}%
\providecommand \BibitemShut  [1]{\csname bibitem#1\endcsname}%
\let\auto@bib@innerbib\@empty
\bibitem [{\citenamefont {Schmidt}\ and\ \citenamefont
  {Lemeshko}(2015)}]{SchmidtLem15}%
  \BibitemOpen
  \bibfield  {author} {\bibinfo {author} {\bibfnamefont {R.}~\bibnamefont
  {Schmidt}}\ and\ \bibinfo {author} {\bibfnamefont {M.}~\bibnamefont
  {Lemeshko}},\ }\href@noop {} {\bibfield  {journal} {\bibinfo  {journal}
  {Physical review letters}\ }\textbf {\bibinfo {volume} {114}},\ \bibinfo
  {pages} {203001} (\bibinfo {year} {2015})}\BibitemShut {NoStop}%
\bibitem [{\citenamefont {Schmidt}\ and\ \citenamefont
  {Lemeshko}(2016)}]{SchmidtLem16}%
  \BibitemOpen
  \bibfield  {author} {\bibinfo {author} {\bibfnamefont {R.}~\bibnamefont
  {Schmidt}}\ and\ \bibinfo {author} {\bibfnamefont {M.}~\bibnamefont
  {Lemeshko}},\ }\href@noop {} {\bibfield  {journal} {\bibinfo  {journal}
  {Physical Review X}\ }\textbf {\bibinfo {volume} {6}},\ \bibinfo {pages}
  {011012} (\bibinfo {year} {2016})}\BibitemShut {NoStop}%
\bibitem [{\citenamefont {Lemeshko}\ and\ \citenamefont
  {Schmidt}(2016)}]{LemSchmidtChapter}%
  \BibitemOpen
  \bibfield  {author} {\bibinfo {author} {\bibfnamefont {M.}~\bibnamefont
  {Lemeshko}}\ and\ \bibinfo {author} {\bibfnamefont {R.}~\bibnamefont
  {Schmidt}},\ }\enquote {\bibinfo {title} {Molecular impurities interacting
  with a many-particle environment: from ultracold gases to helium
  nanodroplets},}\ in\ \href@noop {} {\emph {\bibinfo {booktitle} {Low Energy
  and Low Temperature Molecular Scattering}}},\ \bibinfo {editor} {edited by\
  \bibinfo {editor} {\bibfnamefont {A.}~\bibnamefont {Osterwalder}}\ and\
  \bibinfo {editor} {\bibfnamefont {O.}~\bibnamefont {Dulieu}}}\ (\bibinfo
  {publisher} {RSC},\ \bibinfo {year} {2016})\BibitemShut {NoStop}%
\bibitem [{\citenamefont {Appel}(1968)}]{AppelPolarons}%
  \BibitemOpen
  \bibfield  {author} {\bibinfo {author} {\bibfnamefont {J.}~\bibnamefont
  {Appel}},\ }in\ \href@noop {} {\emph {\bibinfo {booktitle} {Solid State
  Physics}}},\ Vol.~\bibinfo {volume} {21},\ \bibinfo {editor} {edited by\
  \bibinfo {editor} {\bibfnamefont {H.}~\bibnamefont {Ehrenreich}}, \bibinfo
  {editor} {\bibfnamefont {F.}~\bibnamefont {Seitz}}, \ and\ \bibinfo {editor}
  {\bibfnamefont {D.}~\bibnamefont {Turnbull}}}\ (\bibinfo  {publisher}
  {Academic, NY},\ \bibinfo {year} {1968})\BibitemShut {NoStop}%
\bibitem [{\citenamefont {Emin}(2013)}]{EminPolarons}%
  \BibitemOpen
  \bibfield  {author} {\bibinfo {author} {\bibfnamefont {D.}~\bibnamefont
  {Emin}},\ }\href@noop {} {\emph {\bibinfo {title} {Polarons}}}\ (\bibinfo
  {publisher} {Cambridge University Press},\ \bibinfo {year}
  {2013})\BibitemShut {NoStop}%
\bibitem [{\citenamefont {Kuper}\ and\ \citenamefont
  {Whitfield}(1962)}]{PolaronsExcitons}%
  \BibitemOpen
  \bibinfo {editor} {\bibfnamefont {C.}~\bibnamefont {Kuper}}\ and\ \bibinfo
  {editor} {\bibfnamefont {G.~D.}\ \bibnamefont {Whitfield}},\ eds.,\
  \href@noop {} {\emph {\bibinfo {title} {Polarons and Excitons}}}\ (\bibinfo
  {publisher} {Plenum Press, NY},\ \bibinfo {year} {1962})\BibitemShut
  {NoStop}%
\bibitem [{\citenamefont {Devreese}(2007)}]{polaron_review1}%
  \BibitemOpen
  \bibfield  {author} {\bibinfo {author} {\bibfnamefont {J.}~\bibnamefont
  {Devreese}},\ }\href@noop {} {\bibfield  {journal} {\bibinfo  {journal}
  {Journal of Physics: Condensed Matter}\ }\textbf {\bibinfo {volume} {19}},\
  \bibinfo {pages} {255201} (\bibinfo {year} {2007})}\BibitemShut {NoStop}%
\bibitem [{\citenamefont {Devreese}(2015)}]{Devreese15}%
  \BibitemOpen
  \bibfield  {author} {\bibinfo {author} {\bibfnamefont {J.~T.}\ \bibnamefont
  {Devreese}},\ }\href@noop {} {\bibfield  {journal} {\bibinfo  {journal}
  {arXiv:1012.4576}\ } (\bibinfo {year} {2015})}\BibitemShut {NoStop}%
\bibitem [{\citenamefont {Toennies}\ and\ \citenamefont
  {Vilesov}(2004)}]{ToenniesVilesov04}%
  \BibitemOpen
  \bibfield  {author} {\bibinfo {author} {\bibfnamefont {J.~P.}\ \bibnamefont
  {Toennies}}\ and\ \bibinfo {author} {\bibfnamefont {A.~F.}\ \bibnamefont
  {Vilesov}},\ }\href@noop {} {\bibfield  {journal} {\bibinfo  {journal}
  {Angewandte Chemie International Edition}\ }\textbf {\bibinfo {volume}
  {43}},\ \bibinfo {pages} {2622} (\bibinfo {year} {2004})}\BibitemShut
  {NoStop}%
\bibitem [{\citenamefont {Karshenboim}(2005)}]{Karshenboim2005}%
  \BibitemOpen
  \bibfield  {author} {\bibinfo {author} {\bibfnamefont {S.~G.}\ \bibnamefont
  {Karshenboim}},\ }\href@noop {} {\bibfield  {journal} {\bibinfo  {journal}
  {Phys. Rep.}\ }\textbf {\bibinfo {volume} {422}},\ \bibinfo {pages} {1}
  (\bibinfo {year} {2005})}\BibitemShut {NoStop}%
\bibitem [{\citenamefont {Midya}\ \emph {et~al.}(2016)\citenamefont {Midya},
  \citenamefont {Tomza}, \citenamefont {Schmidt},\ and\ \citenamefont
  {Lemeshko}}]{Bikash16}%
  \BibitemOpen
  \bibfield  {author} {\bibinfo {author} {\bibfnamefont {B.}~\bibnamefont
  {Midya}}, \bibinfo {author} {\bibfnamefont {M.}~\bibnamefont {Tomza}},
  \bibinfo {author} {\bibfnamefont {R.}~\bibnamefont {Schmidt}}, \ and\
  \bibinfo {author} {\bibfnamefont {M.}~\bibnamefont {Lemeshko}},\ }\href@noop
  {} {\bibfield  {journal} {\bibinfo  {journal} {arXiv:1607.06092}\ } (\bibinfo
  {year} {2016})}\BibitemShut {NoStop}%
\bibitem [{\citenamefont {Townes}\ and\ \citenamefont
  {Schawlow}(1975)}]{TownesSchawlow}%
  \BibitemOpen
  \bibfield  {author} {\bibinfo {author} {\bibfnamefont {C.~H.}\ \bibnamefont
  {Townes}}\ and\ \bibinfo {author} {\bibfnamefont {A.~L.}\ \bibnamefont
  {Schawlow}},\ }\href@noop {} {\emph {\bibinfo {title} {Microwave
  Spectroscopy}}}\ (\bibinfo  {publisher} {Dover, New York},\ \bibinfo {year}
  {1975})\BibitemShut {NoStop}%
\bibitem [{\citenamefont {Bernath}(2005)}]{BernathBook}%
  \BibitemOpen
  \bibfield  {author} {\bibinfo {author} {\bibfnamefont {P.~F.}\ \bibnamefont
  {Bernath}},\ }\href@noop {} {\emph {\bibinfo {title} {Spectra of atoms and
  molecules}}},\ \bibinfo {edition} {2nd}\ ed.\ (\bibinfo  {publisher} {Oxford
  University Press},\ \bibinfo {year} {2005})\BibitemShut {NoStop}%
\bibitem [{\citenamefont {Varshalovich}\ \emph {et~al.}(1988)\citenamefont
  {Varshalovich}, \citenamefont {Moskalev},\ and\ \citenamefont
  {Khersonskii}}]{Varshalovich}%
  \BibitemOpen
  \bibfield  {author} {\bibinfo {author} {\bibfnamefont {D.~A.}\ \bibnamefont
  {Varshalovich}}, \bibinfo {author} {\bibfnamefont {A.}~\bibnamefont
  {Moskalev}}, \ and\ \bibinfo {author} {\bibfnamefont {V.}~\bibnamefont
  {Khersonskii}},\ }\href@noop {} {\emph {\bibinfo {title} {Quantum theory of
  angular momentum}}}\ (\bibinfo  {publisher} {World Scientific},\ \bibinfo
  {year} {1988})\BibitemShut {NoStop}%
\bibitem [{\citenamefont {Rost}\ \emph {et~al.}(1992)\citenamefont {Rost},
  \citenamefont {Griffin}, \citenamefont {Friedrich},\ and\ \citenamefont
  {Herschbach}}]{RostPRL92}%
  \BibitemOpen
  \bibfield  {author} {\bibinfo {author} {\bibfnamefont {J.~M.}\ \bibnamefont
  {Rost}}, \bibinfo {author} {\bibfnamefont {J.~C.}\ \bibnamefont {Griffin}},
  \bibinfo {author} {\bibfnamefont {B.}~\bibnamefont {Friedrich}}, \ and\
  \bibinfo {author} {\bibfnamefont {D.~R.}\ \bibnamefont {Herschbach}},\
  }\href@noop {} {\bibfield  {journal} {\bibinfo  {journal} {Phys. Rev. Lett.}\
  }\textbf {\bibinfo {volume} {68}},\ \bibinfo {pages} {1299} (\bibinfo {year}
  {1992})}\BibitemShut {NoStop}%
\bibitem [{\citenamefont {Friedrich}\ \emph {et~al.}(1994)\citenamefont
  {Friedrich}, \citenamefont {Slenczka},\ and\ \citenamefont
  {Herschbach}}]{FriedrichCanJPhys94}%
  \BibitemOpen
  \bibfield  {author} {\bibinfo {author} {\bibfnamefont {B.}~\bibnamefont
  {Friedrich}}, \bibinfo {author} {\bibfnamefont {A.}~\bibnamefont {Slenczka}},
  \ and\ \bibinfo {author} {\bibfnamefont {D.}~\bibnamefont {Herschbach}},\
  }\href@noop {} {\bibfield  {journal} {\bibinfo  {journal} {Can. J. Phys.}\
  }\textbf {\bibinfo {volume} {72}},\ \bibinfo {pages} {897} (\bibinfo {year}
  {1994})}\BibitemShut {NoStop}%
\bibitem [{\citenamefont {Lemeshko}\ \emph
  {et~al.}(2011{\natexlab{a}})\citenamefont {Lemeshko}, \citenamefont
  {Mustafa}, \citenamefont {Kais},\ and\ \citenamefont {Friedrich}}]{LemPRA11}%
  \BibitemOpen
  \bibfield  {author} {\bibinfo {author} {\bibfnamefont {M.}~\bibnamefont
  {Lemeshko}}, \bibinfo {author} {\bibfnamefont {M.}~\bibnamefont {Mustafa}},
  \bibinfo {author} {\bibfnamefont {S.}~\bibnamefont {Kais}}, \ and\ \bibinfo
  {author} {\bibfnamefont {B.}~\bibnamefont {Friedrich}},\ }\href@noop {}
  {\bibfield  {journal} {\bibinfo  {journal} {Phys. Rev. A}\ }\textbf {\bibinfo
  {volume} {83}},\ \bibinfo {pages} {043415} (\bibinfo {year}
  {2011}{\natexlab{a}})}\BibitemShut {NoStop}%
\bibitem [{\citenamefont {Lemeshko}\ \emph
  {et~al.}(2011{\natexlab{b}})\citenamefont {Lemeshko}, \citenamefont
  {Mustafa}, \citenamefont {Kais},\ and\ \citenamefont {Friedrich}}]{LemNJP11}%
  \BibitemOpen
  \bibfield  {author} {\bibinfo {author} {\bibfnamefont {M.}~\bibnamefont
  {Lemeshko}}, \bibinfo {author} {\bibfnamefont {M.}~\bibnamefont {Mustafa}},
  \bibinfo {author} {\bibfnamefont {S.}~\bibnamefont {Kais}}, \ and\ \bibinfo
  {author} {\bibfnamefont {B.}~\bibnamefont {Friedrich}},\ }\href@noop {}
  {\bibfield  {journal} {\bibinfo  {journal} {New Journal of Physics}\ }\textbf
  {\bibinfo {volume} {13}},\ \bibinfo {pages} {063036} (\bibinfo {year}
  {2011}{\natexlab{b}})}\BibitemShut {NoStop}%
\bibitem [{\citenamefont {Lemeshko}\ \emph {et~al.}(2013)\citenamefont
  {Lemeshko}, \citenamefont {Krems}, \citenamefont {Doyle},\ and\ \citenamefont
  {Kais}}]{LemKreDoyKais13}%
  \BibitemOpen
  \bibfield  {author} {\bibinfo {author} {\bibfnamefont {M.}~\bibnamefont
  {Lemeshko}}, \bibinfo {author} {\bibfnamefont {R.}~\bibnamefont {Krems}},
  \bibinfo {author} {\bibfnamefont {J.}~\bibnamefont {Doyle}}, \ and\ \bibinfo
  {author} {\bibfnamefont {S.}~\bibnamefont {Kais}},\ }\href@noop {} {\bibfield
   {journal} {\bibinfo  {journal} {Mol. Phys.}\ }\textbf {\bibinfo {volume}
  {111}},\ \bibinfo {pages} {1648} (\bibinfo {year} {2013})}\BibitemShut
  {NoStop}%
\bibitem [{\citenamefont {Altland}\ and\ \citenamefont
  {Simons}(2010)}]{AltlandSimons}%
  \BibitemOpen
  \bibfield  {author} {\bibinfo {author} {\bibfnamefont {A.}~\bibnamefont
  {Altland}}\ and\ \bibinfo {author} {\bibfnamefont {B.}~\bibnamefont
  {Simons}},\ }\href@noop {} {\emph {\bibinfo {title} {Condensed Matter Field
  Theory}}},\ \bibinfo {edition} {2nd}\ ed.\ (\bibinfo  {publisher} {Cambridge
  University Press},\ \bibinfo {year} {2010})\BibitemShut {NoStop}%
\bibitem [{\citenamefont {Pitaevskii}\ and\ \citenamefont
  {Stringari}(2003)}]{Pitaevskii2003}%
  \BibitemOpen
  \bibfield  {author} {\bibinfo {author} {\bibfnamefont {L.~P.}\ \bibnamefont
  {Pitaevskii}}\ and\ \bibinfo {author} {\bibfnamefont {S.}~\bibnamefont
  {Stringari}},\ }\href@noop {} {\emph {\bibinfo {title} {Bose-Einstein
  Condensation}}}\ (\bibinfo  {publisher} {Oxford: Clarendon},\ \bibinfo {year}
  {2003})\BibitemShut {NoStop}%
\bibitem [{\citenamefont {Stone}(2013)}]{StoneBook13}%
  \BibitemOpen
  \bibfield  {author} {\bibinfo {author} {\bibfnamefont {A.}~\bibnamefont
  {Stone}},\ }\href@noop {} {\emph {\bibinfo {title} {The Theory of
  Intermolecular Forces}}}\ (\bibinfo  {publisher} {Oxford University Press},\
  \bibinfo {year} {2013})\BibitemShut {NoStop}%
\bibitem [{\citenamefont {Donnelly}\ and\ \citenamefont
  {Barenghi}(1998)}]{DonnellyHe98}%
  \BibitemOpen
  \bibfield  {author} {\bibinfo {author} {\bibfnamefont {R.~J.}\ \bibnamefont
  {Donnelly}}\ and\ \bibinfo {author} {\bibfnamefont {C.~F.}\ \bibnamefont
  {Barenghi}},\ }\href@noop {} {\bibfield  {journal} {\bibinfo  {journal} {J.
  Phys. Chem. Ref. Data}\ }\textbf {\bibinfo {volume} {27}} (\bibinfo {year}
  {1998})}\BibitemShut {NoStop}%
\bibitem [{\citenamefont {Pentlehner}\ \emph {et~al.}(2013)\citenamefont
  {Pentlehner}, \citenamefont {Nielsen}, \citenamefont {Christiansen},
  \citenamefont {Slenczka},\ and\ \citenamefont
  {Stapelfeldt}}]{PentlehnerPRA13}%
  \BibitemOpen
  \bibfield  {author} {\bibinfo {author} {\bibfnamefont {D.}~\bibnamefont
  {Pentlehner}}, \bibinfo {author} {\bibfnamefont {J.~H.}\ \bibnamefont
  {Nielsen}}, \bibinfo {author} {\bibfnamefont {L.}~\bibnamefont
  {Christiansen}}, \bibinfo {author} {\bibfnamefont {A.}~\bibnamefont
  {Slenczka}}, \ and\ \bibinfo {author} {\bibfnamefont {H.}~\bibnamefont
  {Stapelfeldt}},\ }\href@noop {} {\bibfield  {journal} {\bibinfo  {journal}
  {Physical Review A}\ }\textbf {\bibinfo {volume} {87}},\ \bibinfo {pages}
  {063401} (\bibinfo {year} {2013})}\BibitemShut {NoStop}%
\bibitem [{\citenamefont {Morrison}\ \emph {et~al.}(2013)\citenamefont
  {Morrison}, \citenamefont {Raston},\ and\ \citenamefont
  {Douberly}}]{MorrisonJPCA13}%
  \BibitemOpen
  \bibfield  {author} {\bibinfo {author} {\bibfnamefont {A.~M.}\ \bibnamefont
  {Morrison}}, \bibinfo {author} {\bibfnamefont {P.~L.}\ \bibnamefont
  {Raston}}, \ and\ \bibinfo {author} {\bibfnamefont {G.~E.}\ \bibnamefont
  {Douberly}},\ }\href@noop {} {\bibfield  {journal} {\bibinfo  {journal} {J.
  Phys. Chem. A}\ }\textbf {\bibinfo {volume} {117}},\ \bibinfo {pages} {11640}
  (\bibinfo {year} {2013})}\BibitemShut {NoStop}%
\bibitem [{\citenamefont {Lefebvre-Brion}\ and\ \citenamefont
  {Field}(2004)}]{LevebvreBrionField2}%
  \BibitemOpen
  \bibfield  {author} {\bibinfo {author} {\bibfnamefont {H.}~\bibnamefont
  {Lefebvre-Brion}}\ and\ \bibinfo {author} {\bibfnamefont {R.~W.}\
  \bibnamefont {Field}},\ }\href@noop {} {\emph {\bibinfo {title} {The Spectra
  and Dynamics of Diatomic Molecules}}}\ (\bibinfo  {publisher} {Elsevier, New
  York},\ \bibinfo {year} {2004})\BibitemShut {NoStop}%
\bibitem [{\citenamefont {Lemeshko}\ and\ \citenamefont
  {Friedrich}(2009{\natexlab{a}})}]{LemFriPRArapid09}%
  \BibitemOpen
  \bibfield  {author} {\bibinfo {author} {\bibfnamefont {M.}~\bibnamefont
  {Lemeshko}}\ and\ \bibinfo {author} {\bibfnamefont {B.}~\bibnamefont
  {Friedrich}},\ }\href {\doibase 10.1103/PhysRevA.79.050501} {\bibfield
  {journal} {\bibinfo  {journal} {Phys. Rev. A}\ }\textbf {\bibinfo {volume}
  {79}},\ \bibinfo {pages} {050501} (\bibinfo {year}
  {2009}{\natexlab{a}})}\BibitemShut {NoStop}%
\bibitem [{\citenamefont {Lemeshko}\ and\ \citenamefont
  {Friedrich}(2009{\natexlab{b}})}]{LemFriPRL09}%
  \BibitemOpen
  \bibfield  {author} {\bibinfo {author} {\bibfnamefont {M.}~\bibnamefont
  {Lemeshko}}\ and\ \bibinfo {author} {\bibfnamefont {B.}~\bibnamefont
  {Friedrich}},\ }\href {\doibase 10.1103/PhysRevLett.103.053003} {\bibfield
  {journal} {\bibinfo  {journal} {Phys. Rev. Lett.}\ }\textbf {\bibinfo
  {volume} {103}},\ \bibinfo {pages} {053003} (\bibinfo {year}
  {2009}{\natexlab{b}})}\BibitemShut {NoStop}%
\bibitem [{\citenamefont {Lemeshko}\ and\ \citenamefont
  {Friedrich}(2010)}]{LemFriJPCA10}%
  \BibitemOpen
  \bibfield  {author} {\bibinfo {author} {\bibfnamefont {M.}~\bibnamefont
  {Lemeshko}}\ and\ \bibinfo {author} {\bibfnamefont {B.}~\bibnamefont
  {Friedrich}},\ }\href@noop {} {\bibfield  {journal} {\bibinfo  {journal} {J.
  Phys. Chem. A}\ }\textbf {\bibinfo {volume} {114}},\ \bibinfo {pages} {9848}
  (\bibinfo {year} {2010})}\BibitemShut {NoStop}%
\end{thebibliography}%
\end{document}